\documentclass[aps,superscriptaddress,nofootinbib]{revtex4}

\makeatletter

\usepackage{natbib}
\usepackage{dcolumn}
\usepackage{graphicx}
\usepackage{amsmath}
\usepackage[hang,small,bf,centerlast]{caption} 
\usepackage{rotating}
\begin{document}
\title{Community characterization of heterogeneous complex systems}

\author{Michele Tumminello}
\affiliation{Department of Social and Decision Sciences, Carnegie Mellon University, Pittsburgh, PA 15213, USA.}
\affiliation{Dipartimento di Fisica e Tecnologie Relative, Universit\`a di Palermo, viale delle Scienze Ed.18, I-90128 Palermo, Italia}

\author{Salvatore Miccich\`{e}}
\affiliation{Dipartimento di Fisica e Tecnologie Relative, Universit\`a di Palermo, viale delle Scienze Ed.18, I-90128 Palermo, Italia}

\author{Fabrizio Lillo}
\affiliation{Dipartimento di Fisica e Tecnologie Relative, Universit\`a di Palermo, viale delle Scienze Ed.18, I-90128 Palermo, Italia}
\affiliation{Santa Fe Institute, 1399 Hyde Park Road, Santa Fe NM 87501, USA}

\author{Jan Varho}
\affiliation{Turku Centre for Quantum Physics, Department of Physics and Astronomy, University of Turku, FI-20014 Turun yliopisto, Finland}

\author{Jyrki Piilo}
\affiliation{Turku Centre for Quantum Physics, Department of Physics and Astronomy, University of Turku, FI-20014 Turun yliopisto, Finland}

\author{Rosario N. Mantegna}
\affiliation{Dipartimento di Fisica e Tecnologie Relative, Universit\`a di Palermo, viale delle Scienze Ed.18, I-90128 Palermo, Italia}

\begin{abstract}
We introduce an analytical statistical method to characterize the communities detected in heterogeneous complex systems. By posing a suitable null hypothesis, our method makes use of the hypergeometric distribution to assess the probability that a given property is over-expressed in the elements of a community with respect to all the elements of the investigated set. 
We apply our method to two specific complex networks, namely a network of world movies and a network of physics preprints. The characterization of the elements and of the communities is done in terms of languages and countries for the movie network and of journals and subject categories for papers. We find that our method is able to characterize clearly the identified communities. Moreover our method works well both for large and for small communities.
\end{abstract}
\maketitle 

\section{Introduction}
One of the most important aspects of many complex systems is their organization in clusters or communities. Communities are subsets of the whole system that contain elements with similar features or that are composed of elements strongly interconnected among them. The problem of identifying communities in complex systems is quite complicated and far from being solved in the general case. This problem has been investigated deeply in the case of systems that can be represented by networks.  
Network theory and modeling is today a primary multidisciplinary area of research receiving contributions from many research areas \cite{Watts1998,barabasi1999,newman2002,song2005,schweitzer2009}. 
Statistical physics is one of the main disciplines involved in this effort \cite{Albert2002,Newman2003}. In the past decade, significant progresses have been made to address the problem of detecting communities in complex networks, and many methods have been devised to accomplish this goal (for a recent review on this topic see Ref. \cite{Fortunato2010}).

However the identification of communities is only the first step toward the understanding of the structure of a complex system. The natural next step is to interpret  what these communities actually represent. In other words after the communities have been identified, it is important to characterize them in terms of the attributes
shared by the elements belonging to the same community.
As an example consider the case, investigated in detail below, of the set of movies. 
Classes of attributes of a movie can be, for example, the genre, the production country, the filming location, the
language, the movie director, the casting director, and the production company. It is worth noting that multiple attributes of the same class might be associated with the same element of the system. For example a movie might be produced by production companies of different countries.
The question is thus to find a quantitative way to establish whether a given community of movies can be characterized by one or more specific countries, one or more languages, etc.. In this paper we introduce a quantitative statistical method to characterize a community by identifying those attributes that characterize it. These attributes are those that are not statistically consistent with a null hypothesis of random occurrence of the attributes across all the elements of the system. Note that these attributes are not necessarily the most frequent in the community, but rather are those whose frequency in the community is markedly larger than the frequency observed in the whole set.

Two important comments are in order. First, since many community detection algorithms present statistical fluctuations and the elements of the system may be very heterogeneous, we should not expect that the elements of any empirically identified community  turn out to be fully homogeneous. This observation suggests that the characterization of communities is often a complicated task.
Our method for the characterization of communities is also of statistical nature and takes properly into account the compositional heterogeneity of the detected communities.  
The second comment is that our method is not a community detection method but rather it can be applied to the output of any community detection algorithm. Our method is very general and can be applied to systems with different structural properties (i.e. not necessarily networks). In this paper we will show the 
effectiveness of the method in the characterization of communities detected in the projected network of a bipartite complex system. Bipartite networks are composed by two disjoint sets of nodes such that every link connects a node in the first set with a node of the second set. The bipartite network is often transformed by one-mode projecting, i.e. one creates a network of nodes belonging to one of the two sets and two nodes are connected when they have at least one common neighboring node of the other set.  Classic examples of bipartite systems are movies and actors \cite{Watts1998,barabasi1999,song2005}, authors and scientific papers \cite{newman2001,barabasi2002,guimera2005,colizza2006}, email accounts and emails \cite{McCallum2007}, mobile phones and phone calls \cite{Onnela2007}, plants and animals that pollinate them \cite{bascompte2003, reed2009}. Here we present and discuss our method by investigating two classic networks widely studied in the network literature. The first is the movie network obtained as a projected network from the bipartite relationship between actors and movies where the actors play a role. The second is the network of papers obtained projecting the relationship between authors and papers written by them. 

The paper is organized as follows. In Section II we present the two bipartite systems used to illustrate the application of our quantitative method for the characterization of communities. In Section III we introduce our method and in Section IV we present the results obtained from the analysis of the communities in the two bipartite systems. Section V concludes.

\section{The investigated systems}

We investigate two systems naturally represented by bipartite networks, namely the movie-actor system and the paper-author system. From these bipartite systems we construct the projected network of movies and papers, respectively. Projected networks might keep different level of information about the nature of their links. The simplest level of information is about the existence or non-existence of the link. In this case the network is addressed as adjacency network. More informative networks are networks where links can be weighted \cite{Newman2001b}  or even statistically validated \cite{Tumminello2010svn} to take into account the heterogeneity of elements of the systems. 

In the present paper we propose a method to characterize the partitions of a network, therefore our focus is neither on the way of constructing projected networks nor on a new partitioning method. For this reason and for the sake of simplicity here we construct and then partition the network in the most elementary and direct way. Specifically we investigate the adjacency network and we use the Infomap method by Rosvall and Bergstrom \cite{Rosvall2008} to detect communities. This algorithm is considered one of the best algorithms \cite{Fortunato2010} of community detection in networks. The method uses the probability flow of random walks  to identify the community structure of the system in the considered network. This approach implies that two independent applications of the method to the same network may produce (typically slightly) different partitions of vertices. For each investigated network, we run the Infomap $10^3$ times and we select the best partition according to the minimal ``code length" \cite{Rosvall2008}.

\subsection{The movie-actor system}
The investigated set of movies and actors is obtained from the Internet Movie Database (IMDb) \cite{IMDb}, which is the largest web repository of world movies. We consider here the bipartite relationship between movies and actors for  22,423  movies classified within the genre Drama produced during the period 1990-2008 all over the world\footnote{Each movie can also be classified by additional genres beside Drama. We only exclude from the set the movies which are also classified within the genre Short to ensure similar planning of the cast.}. We choose to focus on a single genre to have a relatively homogeneous set of movies and to avoid to consider movies where actors are not professionals like in movies of genre Documentary. The set includes movies realized in 160 countries and the number of involved actors is 197,858. From the bipartite relationship between actors and movies we construct the adjacency network of movies by connecting with a link two movies when at least one actor played in both of them. 

There are 1,222 isolated nodes corresponding to movies with actors that did not act in any other movies of our set. We restrict here our attention to the set of movies with at least one link. By excluding the isolated nodes, our set is composed by 21,201 movies\footnote{The results do not change significantly if we include also the isolated nodes.}. The adjacency network is naturally divided in 63 connected components. The largest 
one comprises 21,063 movies whereas the second largest component has only 6 movies. This shows that the movie industry all over the world is highly connected in a giant component and actors of different countries set up joint projects. The number of links of the network is 656,609 indicating a high connectivity among movies (the average degree of movies is 31.0).  

When we apply the Infomap partitioning algorithm to the adjacency network of movies, we obtain a partitioning of the network which presents 369 distinct clusters. The cluster size decreases from the largest value of 5,048 movies smoothly down to the smallest value of 2. The largest eleven clusters have more than 500 movies each and there are 132 clusters composed by just two movies. The average cluster size is 57.4 and its standard deviation 296 reflecting a wide range of cluster size. 

A movie is characterized by many classes of attributes, e.g. the genre, the production country, the filming locations, the languages, the movie director, the casting director and the production company. In the present study we choose to characterize movies with respect to the production countries and to the language spoken in it\footnote{In the IMDb database a movie might be associated with more than one language. We consider all the languages associated to a movie.}.  This choice is subjective and should not be considered as exhaustive of the cluster characterization. In fact our method provides a tool to detect over-representation of any characteristic selected by the investigator. Each specific choice does not guarantee a complete specification of the characteristics of investigated clusters but rather provides a statistically robust evidence of the role or of the absence of the role of the selected attributes.

\subsection{The paper-author system} 

As a second example, we consider the bipartite system of scientific papers and authors. We collect information about all the 31,768 pre-print papers posted in the physics archive [21] during the time period 2000-2009. Distinct authors are identified by their family name and by all their initials. From the bipartite relationship between authors and papers we construct the adjacency network of papers by connecting with a link two papers when at least one author is present in both of them. 

There are 2,898 isolated nodes of papers written by authors of only one paper in our set.  As before we consider only the 28,879 papers represented by non isolated nodes. The largest component of the adjacency network of papers comprises 20,676 papers, whereas the second largest component has just 118 papers. Again the international research community of physicists posting their manuscripts into the physics archive from all over the world is highly connected and authors of different countries publish joint papers. The number of links of the network is 272,562 indicating a quite high average connectivity (the average degree of papers is 9.4), although smaller than the one observed for the network of movies.

When we apply the Infomap partitioning algorithm to the adjacency network of papers we obtain a partitioning of the network which presents 5,746 distinct clusters. The cluster size decreases from the largest value of 246 papers smoothly down to the smallest value of 2. The largest fifteen clusters have more than 100 papers and there are 785 clusters composed by just two papers. Also for this system we observe a wide range of cluster size.

A paper is characterized by many classes of attributes, e.g. the subject category, the nationality of authors, the affiliation of authors and the journal where the paper is eventually published. Here we focus on two of them. Specifically, we characterize papers with respect to the journal where the paper has been eventually published (we use the Journal-ref field of the record) and with respect to the archive classification of the subject of the research area (including cross-referencing). Again this choice is done just for illustrative purposes of the proposed method and it should not be considered as an exhaustive characterization of the system. Note that in the case of papers the information on the journal reference is not always available. In our set of 28,879 papers there are 16,397 papers without a reference. Moreover for 3,784 papers the reference is written in a non standard way, i.e. we were not able to automatically identify the corresponding journal. For these two cases we consider that the journal attribute of the paper is ``Not Available" and ``Not Standard", respectively, and we test their occurrence in the different clusters. 

\section{Over-expression of characterizing properties}

We now present our method for characterizing the clusters of a partition in terms of the attributes of the elements belonging to the cluster. Our method is inspired by the procedure of statistical validation of over-expression of genes belonging to a list of interest in specific terms of the Gene Ontology database  \cite{Draghici2003}. The Gene Ontology \cite{GOC} is a controlled vocabulary where sets of genes and gene attributes are organized in a standardized way in groups called terms, which are associated with specific biological processes, molecular functions and cellular components. 

Here we adapt the procedure used in the statistical validation of over-expression of genes in specific terms of the Gene Ontology to the problem of estimating over-expression of elements attributes in a specific cluster of a projected network. In our investigation, we label each element with selected attributes. For example a given movie may turn out to be produced in countries $N_X$, $N_Y$, and $N_Z$ and spoken in languages  $L_X$ and $L_Y$. 
Let us now consider a system of $N$ elements and a specific cluster $C$ of $N_C$ elements to be characterized.
A schematic illustration of the investigated co-occurrence is given in Fig. 1. Each element of the system has a certain number of attributes belonging to a specific class, e.g. production countries. We indicate the total number of different attributes over all the elements of the system with $N_A$ (in our example the total number of production countries in the set). For each attribute $Q$ of the system, e.g. the USA as production country, we test if $Q$ is over-expressed in cluster $C$. In other words, we test if the number $N_{C,Q}$ of elements in cluster $C$ that have the attribute $Q$ is significantly larger than what expected by randomly selecting the $N_C$ elements of the cluster from the total $N$ elements of the system. The probability that $X$ elements in cluster $C$ have the attribute $Q$, under the null hypothesis that elements in the cluster are randomly selected, is given by the hypergeometric distribution \cite{Feller}
\begin{equation}
H(X|N,N_C,N_Q)=\frac{{N_C \choose X} {N-N_C \choose N_Q-X}}{{N \choose N_Q}}
\label{hyper}
\end{equation}
where $N_Q$ is the total number of elements in the system with attribute $Q$. 
By using this distribution one can associate a $p$-value to the observed number $N_{C,Q}$ of elements in cluster $C$ that are classified with the attribute $Q$ according to the equation
\begin{equation}
\label{cluschar}
p(N_{C,Q}) =1- \sum_{X=0}^{N_{C,Q}-1} H(X|N,N_C,N_Q).
\end{equation}
If $p(N_{C,Q})$ is smaller than a given statistical threshold $p_t$ (we use $p_t=1\%$ throughout this paper) we say that the attribute $Q$ is over-expressed in cluster $C$ and therefore the attribute $Q$ is a characterizing attribute of cluster $C$. We separately test all the possible $N_A$ attributes for each detected cluster $C$. Therefore our test is a multiple hypothesis test and the statistical threshold must be corrected. The most stringent method to adjust the statistical threshold for multiple hypothesis test is to use the Bonferroni correction \cite{Miller1981}. This correction is based on the consideration that if one tests $N_t$ either dependent or independent hypotheses on a set of data, then a conservative way of maintaining the error rate low is to test each individual hypothesis at a statistical significance level of $p_t/N_t$. In our investigations, for each class of attributes we use a statistical threshold $p_b$ corrected for multiple hypothesis testing by using the Bonferroni correction, i.e. we set $p_b=0.01/N_A$.
\begin{figure}
\includegraphics[width=0.5\textwidth]{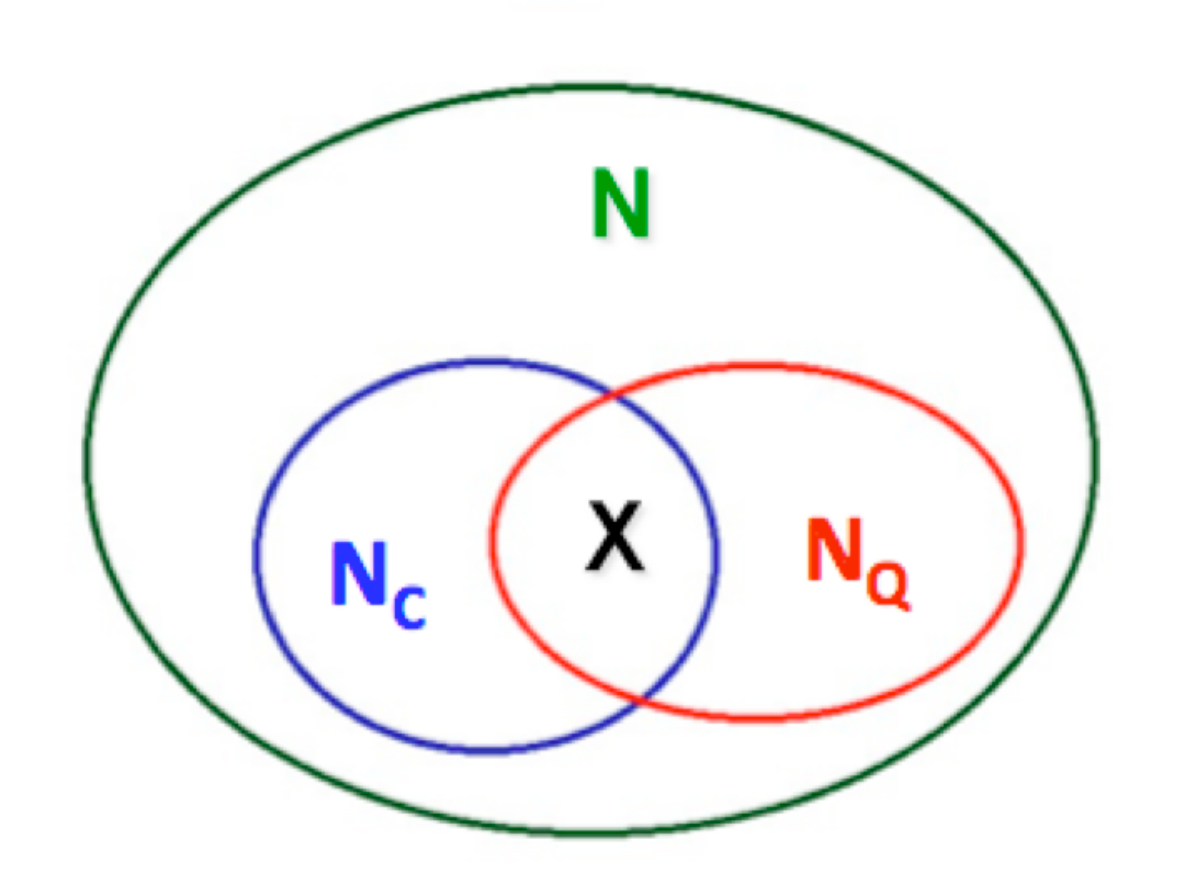}
\label{fig1}
\caption{Illustrative scheme of the co-occurrence investigation quantitatively evaluated by Eq. \ref{hyper}. $N$ is the number of elements of the system. $N_C$ is the number of elements in cluster $C$. $N_Q$ is the number of elements with a specific attribute $Q$ (for example USA as production country) and $X$ is the number of elements in cluster C having Q among its attributes.}
\end{figure} 

\section{Characterization of clusters}

\subsection{The movie network}
In the case of the movie network for each identified cluster (obtained as described in Section II) we test the over-expression of each production country and, separately, of each language. Table~\ref{adjmovies} shows results obtained for the ten largest clusters and for ten medium sized clusters. These are the largest clusters of size smaller than 50, which is a size two orders of magnitude smaller than the size of the largest cluster (5,048). We include in our analysis these relatively small clusters to test the effectiveness of our method for different cluster sizes. For each cluster the table reports the over-expressed countries and the over-expressed languages.

An analysis of the over-expressed countries and languages results to be highly informative. 
For example the largest cluster is characterized by USA as production country and English and American Sign Language as languages. It is important to stress that, while English is the most common language in the cluster (it is the language in 4,888 movies of the 5,048 ones in the cluster), the American Sign Language is quite rare in the cluster being used only in 13 movies. Languages with higher occurrences in the cluster are: Spanish (203), French (116), German (65), Italian (56), Russian (48), Mandarin (35), Arabic (28), Japanese (27), Hebrew (17), Cantonese (17), Korean (17), Portuguese (14) and Vietnamese	(13). However in the whole investigated set the American Sign Language occurs only in 14 movies. The statistical validation of the over-expression is due to the fact that 13 out of the 14 movies are present in this cluster and such occurrence cannot be explained by the random null hypothesis according  to Eq. (1) and Eq. (2). 
Conversely, the occurrence of the remaining 73 languages represented in the cluster is compatible with the random null hypothesis taking into account their frequency heterogeneity.  
This example shows that our method is able to identify the over-expressed attributes in a cluster with respect to the whole investigated set. These attributes might be quite unfrequent in absolute terms (even in the cluster), as in the American Sign Language case.

A comparison of over-expressed countries and languages from Table~\ref{adjmovies} indicates a meaningful relation between countries and languages spoken in them for several clusters.
For example, let us consider cluster of rank 2 where Philippines is over-expressed as production country. In this cluster the over-expressed languages are Filipino, Tagalog and Visayan which are all languages of the Philippines. Similarly, in cluster of rank 4, India is over-expressed and the eight over-expressed languages are all primarily spoken in this country (Assamese, Bengali, Bhojpuri, Gujarati, Hindi, Marathi, Punjabi and Urdu). Another representative case is observed in cluster of rank 8. This cluster is characterized by the German language and by its regional variance in Switzerland (Swiss German) and the over-expressed production countries are all speaking this language (Austria, East Germany, West Germany, Germany and Switzerland).

In a few cases the parallel comparison of country and language is not straightforward like in the cases of cluster of rank 3 and cluster of rank 5, where some over-expressed production countries  (like Germany in cluster of rank 3) have not a correspondent language over-expression. The reason for that could be that in these cases our method provides false positive or the unexpected over-expressions might be due to the specific structure of the obtained partition which might be suboptimal. In fact, we have preliminary evidence that the over-expression characterization of network partitions improves when the partitioning is obtained in weighted and in statistically validated networks \cite{Tumminello2010svn}.

The method also provides meaningful results in relatively small clusters. In Table~\ref{adjmovies} the ten clusters of size ranging from 43 to 26 elements are all showing a meaningful characterization of production country and spoken language.

\subsection{The paper network}

In a similar way we perform the analysis of the paper network.
In Table~\ref{adjpapers} we show the results obtained for the paper network when papers posted into the pre-print archive are characterized by the subject category and by the journal where the paper was eventually published (when available). Similarly to what we have done for the movie network, we show the ten largest clusters and 18 medium size clusters. These 18 clusters are all clusters of size 24 which is the largest integer number one order of magnitude smaller than the size of the largest cluster (246).

The over-expression of a subject category characterizes research areas of papers belonging to the considered cluster. A parallel analysis of journal and subject category over-expression provides meaningful characterizations. For example, in the cluster of rank 4 of Table~\ref{adjpapers}, papers presenting over-expression of fluid dynamics (phy.flu-dyn), chaotic dynamics (nlin.CD), and cellular automata lattice gas (nlin.CG) also present a parallel over-expression of journals: Journal of Fluids Mechanics, Physics of Fluids and Physica D. It is worth noting that 16 different journals are present in this cluster. Moreover the occurrence of Journal of Fluids Mechanics, Physics of Fluids, and  Physica D is only the 5th, the 7th and the 8th, respectively, showing again that the method takes into account properly the heterogeneity of the different journals. For this cluster the top three occurrences are ``Not Available" (78, meaning that no information is provided for those papers), PRE (13) and PRL (12). 
Another result clearly showing the effectiveness of our method concerns cluster of rank 7. In this cluster the over-expression involves pattern formation and solitons (nlin.PS), optics (phy.optics), material sciences (cm.mtrl-sci) and condensed matter (cm). This cluster presents a parallel over-expression of the journals PRL, Optics Letters, Opto Express and Applied Physics Letters.

Similar analysis can be performed for most of the clusters and in many cases one recognizes the specialization of over-expressed journals. The method also works well for medium sized clusters. In Table~\ref{adjpapers} clusters specialized in chemical physics (rank 280), astrophysics space research (rank 283), optics (rank 287 and 288), fluid dynamics (rank 294 and 296) and instrumentation and detectors (rank 295 and 297) are easily detected.

\section{Conclusions}

We have introduced a statistical method to characterize the communities of a complex system in terms of the over-expressed attributes found in the elements of the community. Our method tests the actual observations against the null hypothesis that a given attribute is uniformly distributed across the elements of the system. 
By using the Bonfrerroni correction for multiple hypothesis testing, our method results to be very robust with the respect to the appearance of false positives.
This constitutes the first step toward the interpretation of the detected communities. In this paper we have presented the application of our method to the movie network and to the physics preprint network. We have constructed the projected unweighted adjacency networks from the original bipartite system and we have identified the communities by using the Infomap method. The results of the analysis show that both large and medium sized communities are very often clearly characterized in terms of the over-expressed attributes. 

We would like to stress again that our method can be applied to any partition of any system. The investigated system should not necessarily be represented by a network and the method used to partition the system can be arbitrary. The only critical assumption is the uniformity assumption of the null hypothesis. It is however possible, at least in principle, to generalize the method presented here to different null hypothesis. The main difficulty is the computation, either analytically or numerically, of the $p$-value in analogy to what is done in Eq. (\ref{cluschar}).

{\bf Acknowledgements} J.P. acknowledges financial support by the Magnus Ehrnrooth Foundation and  the Vilho, Yrj\"o, and Kalle V\"ais\"al\"a Foundation.

\newpage

\begin{center}
\begin{widetext}
\begin{sidewaystable}
\begin{center}
%\begin{tiny}
\begin{small}
\caption{Movie network. Over-expression of production country and language for the ten largest clusters obtained from the Infomap partitioning of the adjacency unweighted network and for ten medium size clusters. $N_{DC}$ and $N_{DL}$ are the number of distinct countries and languages, respectively, present in the cluster. Production country is indicated with the standard international three letter code. The exceptions are East Germany (EDEU), West Germany (WDEU), and Soviet Union (SovUni). The two numbers in parentheses are the number of occurrences of the attribute in the cluster and in whole set. For example USA (4888/7072) in the first line indicates that in cluster of rank 1 there are 4888 USA movies, while in the whole investigated set there are 7072 USA movies.} \label{adjmovies}
\begin{tabular}{|c|c|c|l|c|l|} \hline
rank  & Size & $N_{DC}$  & Country over-expressions & $N_{DL}$  & Language over-expressions \\  \hline
 1    & 5048      &  68     & USA (4888/7072)  &  75   &  English (4924/9889) American Sign Language (13/14) \\ \hline
 2    & 1033      &  13     & PHL (1012/1018)   &  15   &  Filipino (979/987) Tagalog (984/997) Visayan (9/10) \\ \hline
 3    & 1023      &  46     & DEU (114/1292) IRL (130/156) LUX (16/50) GBR (811/1411)   &  53   &  English (986/9889) Irish Gaelic (11/15) Scottish Gaelic (5/5) \\ \hline
 4    &   994      &  24     & IND (924/1823)                                                                                     &  75   &  Assamese (6/10) Bengali (22/122) Bhojpuri (5/5) Gujarati (9/9) \\
 ~    &        ~      &    ~     & ~                                                                                                               &  ~     &  Hindi (821/876) Marathi (43/43) Punjabi (30/39) Urdu (32/51) \\ \hline
 5    &   909      &  54     & DZA (12/16) BEL (106/247) FRA (837/1772) ITA (64/851)          &  54   &  Arabic (41/271) French (833/1455) \\
 ~    &          ~    &    ~     & LUX (15/50) MAR (12/41) CHE (69/204)                                        &   ~     &  ~ \\ \hline
 6    &   797      &  25     & JPN (782/986)                                                                                      &  22   &  Japanese (785/974) \\ \hline
 7    &   741      &  18     & CHN (258/336) HKG (467/509) SGP(11/44) TWN (79/117)        &  27   &  Cantonese (425/490) Mandarin (351/567) Shanghainese (8/10)  \\
 ~    &        ~      &    ~     & ~                                                                                                              &  ~      &  Thai (14/92) Taiwanese (14/27)  \\ \hline
 8    &   655      &  38     & AUT (80/132) EDEU (5/5) DEU (550/1292) CHE (57/204)          &  36   &  German (595/961) Swiss German (18/22) \\
 ~    &       ~      &     ~     & WDEU (5/11)                                                                                         &  ~     &  ~  \\ \hline
 9    &  578      &   31     & ITA (554/851) FRA (80/1772)                                                             &  29   &  Italian (520/758) \\ \hline
10   & 574      &    42     & BLR (21/22) KAZ (14/20) RUS (398/443) SovUni (108/131)       &  32   &  Chechen (5/5) Kazakh (4/8) Russian (528/728) Ukrainian (12/20) \\
 ~    & ~           &   ~        & UKR (39/42)                                                                                           &  ~     &  ~  \\ 
\hline
\hline
48  & 43          &     9     & CHL (43/63)  &  4   &  Spanish (41/1646) \\ \hline
49  & 41          &     8     & USA (36/7072)   &  2   &  English (37/9889) \\ \hline
50  & 39          &     6     & GEO (36/45)  SovUni (8/131)  &  6   &  Georgian (39/54) \\ \hline
51  & 34          &     1     & THA (34/70)   &  4   &  Thai (34/92) \\ \hline
52  & 31          &     4     & USA (29/7072)   &  5   &  English (31/9889) \\ \hline
53  & 31          &     1     & JPN (31/986)   &  1   &  Japanese (30/974) \\ \hline
54  & 30          &  13     & EST (21/30) FIN (6/127) SovUni (6/131)   &  8   &  Estonian (23/29) Latvian (4/23) Lithuanian (3/17) \\ \hline
55  & 28          &     6     & SGP (25/44)   &  11   &  Cantonese (6/490) Hokkien (17/29) Malay (5/76) Mandarin (17/567) \\ \hline
56  & 27          &     9     & TWN (26/117)   &  10   &  Mandarin (25/567) Taiwanese (11/27) \\ \hline
57  & 26          &     3     & USA (26/7072)   &  4   &  English (25/9889) \\  
 \hline
\end{tabular}
\end{small}
%\end{tiny}
\end{center}
\end{sidewaystable}
\end{widetext}
\end{center}

\newpage

%\begin{center}
\begin{widetext}
\begin{sidewaystable}
\begin{center}
%\begin{tiny}
%\begin{scriptsize}
\begin{small}
\caption{Paper network. Over-expression of journal and subject category for the ten largest clusters obtained from the Infomap partitioning of the adjacency unweighted network of papers and for ten medium size clusters. $N_{DJ}$ and $N_{DS}$ are the number of distinct journals and subject categories, respectively, present in the cluster. The two numbers in parentheses are the number of occurrences of the attribute in the cluster and in whole set. For example PRL (22/970) in the first line indicates that in cluster of rank 1 there are 22 papers published in PRL, while in the whole investigated set there are 970 papers published in PRL.} 
\label{adjpapers}
\begin{tabular}{|c|c|c|l|c|l|} \hline
cluster rank  & Size & $N_{DJ}$ &  Journal over-expressions & $N_{DS}$ &  Subject over-expressions \\  
\hline
 1    & 246      &  18   &  PRL (22/970) PRD (7/88) PRC (5/37) PRA (78/819) &  34   &  nucl-th (41/432) hep-ph (47/747) phy.atom-ph (232/3855) \\ 
 ~    & ~   &  ~ &  JPB (8/139) &  ~ &  astro-ph (32/1184) hep-th (15/540) \\ 
 \hline
 2    & 197  &  24 &  PRA (48/819) EPJD (7/70) CJP (4/21) &  24   &  q-bio.CB (6/81) phy.atom-ph (172/3855)\\ 
 \hline
 3    & 158  &  12 &  PhysPlas (19/122) & 25  &  phy.space-ph (31/675) phy.plasm-ph (127/2063) \\
 \hline
 4    &   153  &  16   &  PhysicaD (6/43) PhysFlu (7/81) JFluMec (10/49) &  38   &  phy.flu-dyn (125/2556) nlin.CD (64/836) \\
 ~    &        ~      &    ~    & ~ &  ~   & nlin.CG (6/52) \\ 
 \hline
 5    &  144 &  20   &  PRE (20/992) PhysicaA (10/305) NJP (6/144) &  42 &  q-bio (8/236) phy.soc-ph (73/2798) cs.IR (5/53) \\
 ~    &          ~    &     ~  & ~ &    ~ & phy.data-an (51/1860) phy.bio-ph (27/1981) \\ 
 \hline
 6    & 133 &  14   &  EConfC (20/332) ConfProc (8/50) &  18   &  phy.acc-ph (119/1249) \\ 
 \hline
 7    & 130  &  14 &  PRL (15/970) OptLett (13/183) OptoExp (15/145) &  23   &  nlin.PS (40/482) phy.optics (119/4252) cm.mtrl-sci (12/634) \\
 ~    &    ~  &    ~   &  AppPhysLett (11/112) &  ~      &  cm (8/273) \\ \hline
 8    & 119  & 14   &  PNAS (11/73) PRE (28/992)  &  34   &  cm.stat-mech (59/1827) q-bio.PE (10/243) phy.soc-ph (101/2798) \\
 ~    &       ~      &     ~    & ~ &  ~  &  cs.NI (17/62) cs.MA (9/34) cs.IR (5/53) cs.HC (5/13)  \\ 
~    &       ~      &      ~  & ~ &  ~   &  cm.dis-nn (32/471) cs.GT (6/19) cs.CY (8/61) \\ 
\hline
 9    &  118  &  9 &  NuoCimB (3/12) Chaos (20/73) &  2   &  phy.gen-ph (118/3700) \\ 
 \hline
10   & 114 & 17 &  --  &  31  &  phy.optic (45/4252) q-fin.ST (22/381) \\
\hline
\hline
280  & 24  &  7  &  JChemPhys (4/227)  &  9   &  cm.mtrl-sci (23/634) phy.chem-ph (23/2226) phy.atom-ph (14/3855) \\ 
\hline
281  & 24  & 5 &  -- &  2   &  phy.class-ph (16/1914) phy.acc-ph (19/1249) \\ 
\hline
282  & 24 & 3 & --  &   5  &  phy.gen-ph (18/3700) \\ 
\hline
283 & 24 & 4 &  AstroPhysSpaceSc (7/24) &  3 & phy.gen-ph (23/3700) \\ 
\hline
284 & 24 & 5 & - -- &  3   &  phy.gen-ph (24/3700) \\ 
\hline
285  & 24 & 9 &  PRE (8/992) &  14 &  q-bio.PE (4/243) phy.soc-ph (20/2798) \\ 
\hline
286  & 24 & 5 &  -- &  17 &  cm.stat-mech (10/1827) phy.soc-ph (15/2798) q-fin.GN (7/193) \\ 
\hline
287  & 24 & 9 &  PRB (4/217) AppPhysLett (4/112)   &  8  &  phy.optics (23/4252) \\ 
\hline
288  & 24 & 8 &  PRA (8/819)  & 4  & phy.optics (23/4252)\\ 
\hline
289  & 24 & 9 & - -- & 13 &  phy.flu-dyn (11/2556) \\  
\hline
290 & 24 & 10 &  -- & 15 & cm.stat-mech (8/1827) nlin.PS (11/482) cm.other (7/589) \\  
~ & ~  &      ~  & ~  &   ~ & nlin.CD (10/836)  \\
\hline
291 & 24 & 3  &  NotStand (17/3784) GeophysResLett (3/17)  &  9  & phy.geo-ph (18/943) \\  
\hline
292 & 24 & 7 &  -- & 7 &  phy.space-ph (22/675) phy.plasm-ph (9/2063) \\ 
\hline
293 & 24 & 5 &  -- &  17  & nlin.PS (7/482) phy.optics (16/4252) \\  
\hline
294  & 24 & 2 &  NotAvail (23/16397) & 15 &  phy.flu-dyn (16/2556) nlin.CD (8/836) phy.ao-ph (6/812) \\  
\hline
295  & 24 & 5   & NuclInstrMeth (8/456) & 14 &  phy.ins-det (14/1843)\\  
\hline
296 & 24 & 5   & PhysFluids (4/81) & 14 &  phy.flu-dyn (16/2556) \\  
\hline
297 & 24 & 5 &  NuclInstrMeth (8/456) & 12 &  phy.ins-det (9/1843) phy.data-an (9/1860) astro-ph (12/1184) \\  \hline

\end{tabular}
%\end{scriptsize}
\end{small}
%\end{tiny}
\end{center}
\end{sidewaystable}
\end{widetext}
%\end{center}

\end{document}